\documentclass[12pt,a4paper,twoside]{article}        %###
\usepackage[russian, english]{babel}                 %###
\usepackage[cp1251]{inputenc}                        %###
\usepackage{mmgart}                                  %###
\def\bb {\begin {eqnarray}}
\def\ee {\end {eqnarray}}

\begin{document}

%################ Do not change this block ##############
%### 1st page, volume-issue numbers, range of pages  ####
\setcounter{page}{27}                                %###
\thispagestyle{empty}                                %###
\begin{heading}                                      %###
{Volume\;2,\, N{o}\;2,\, p.\,27 -- 36\, (2014)}      %###
{}%Special issue                                     %###
\end{heading}                                        %###
%########################################################

%######################### STEP 1 #######################
%####################### Type a title ###################
\begin{Title}
Mathematical Modeling of
Graphite-to-Diamond Transition
\end{Title}
%########################################################

%######################### STEP 2 #######################
%##### Insert authors' names, addresses and e-mails #####
\begin{center}
\Author{1}{E.M.~Sarkisyan},
\Author{1,2,a}{K.B.~Oganesyan},
\Author{1}{N.Sh.~Izmailian}
\Author{2}{and E.A.~Ayryan}
\end{center}

%===================== Addresses
\begin{flushleft}
\Address{1}{Yerevan Physics Institute, Alikhanian Brothers 2, 375036 Yerevan}\\
\Address{2}{Joint Institute for Nuclear Research, Dubna,  Russia}\\
%===================== e-mails
\Email{$^a$bsk@mail.yerphi.am}
\end{flushleft}
%############################################################

%######################### STEP 3 ###########################
%##### Insert the first author's name and a short title #####
\Headers{E.M.~Sarkisyan et al}{
Mathematical Modeling of Graphite-to-Diamond Transition}
%############################################################

%########### Do not change this block ################
\begin{flushleft}                                 %###
\small\it Received 28 July 2014.                  %###
Published 26 August 2014.                         %###
\end{flushleft}                                   %###
%#####################################################

%######################### STEP 4 ###########################
%### Insert thanks if needed, otherwise delete this block ###
\Thanks{One of the authors (E.A.~Ayryan) is grateful for the support
from the Russian Foundation for Basic Research,  grants No. 14-01-00628 and No. 13-01-00595.
}
%############################################################

%################ Do not change this block ##################
%############################################################
\Thanks{\mbox{}\\
\copyright\,The author(s) 2014. \ Published by Tver State University, Tver, Russia}
\renewcommand{\thefootnote}{\arabic{footnote}}
\setcounter{footnote}{0}
%############################################################

%######################### STEP 5 ###########################
%# Type abstract (up to 200 words), key words and MSC/PACS ##
\Abstract{The energetic evaluations of graphite-to-diamond transition by electron irradiation are performed.  The heat conduction problem is solved for
the diamond synthesis when a pulse-periodic source of energy
is located within a   graphite cylinder;
time dependences of temperature and pressure are found.
It is shown, that the temperatures and pressures implemented in graphite are
sufficient for graphite-to-diamond transition under electron bombardment.}


%############################################################

%######### Do not change this block #######
%##########################################
\renewcommand{\baselinestretch}{1.1}   %###
\newpage                               %###
%##########################################

\section{Introduction}
In
%papers
Refs. \cite{Bats1},\cite{Bats2} an opportunity  of
%use
using a
%higher-current
high-current
%pulse
pulsed
relativistic electronic beam (REB) for
%realization of
implementing
structural transformations in graphite, carbides
and
%nitride of boron
boron nitride
is experimentally proved. The opportunity
%of the
to use
%of
pulse-periodic electron accelerators with rather small current has been specified in the offer
Ref. \cite{Am} for the same purposes.

Let us consider an opportunity of
%transformation
transforming
graphite to
diamond using electron irradiation.
One can estimate the
%necessary energy
energy, necessary
for synthesis as follows.
%Graphite
The graphite
density is
$2.21$ g/cm$^3$,
 the diamond density
% --
is $3.51$ g/cm$^3$.
The ratio
%of
equals specific volumes $V_d/V_{gr}=0.63 $, i.e. during
the compression the volume of graphite
is changed
%on
by 37\%.
Starting from the dependence of potential
energy of interaction on its specific volume, one can estimate,
that such compression requires
the  energy $\epsilon=1$eV/atom
%(correspondingly
(the corresponding
pressure $P=10^6$ Atm \cite{Zeld},\cite{Alt}.

The
%energetic
energy
barrier, during
to synthesis of diamond from graphite, amounts
$20\div 30$ Cal/mole (\cite{Coll}).
%note,
Note that in that
%paper
Ref. \cite{Coll}
are
the presented
%quantities
values
of the energy barrier are $50\div 250$ Cal/mole
%of energetic barrier
for
the inverse process of graphitization of diamond. According to other
estimates, the activation energy for the direct
%process graphite to  diamond
graphite-to-diamond process
amounts
%about
to nearly $200$Cal/mole \cite{Kurd}.
%%%%%%%%%%%%%%%%%%%%%%%%%%%%%%%%%%

These estimates are based on
the fact, that during
the graphite-to-diamond
transformation
%graphite to diamond
the type of bond is changed.
For
%graphite to diamond
the graphite-to-diamond
transformation it is necessary to transform three sp$^2$-bonds and one p- bond of graphite
%to
into four sp$^3$- hybrid bonds of tetrahedral oriented diamond in space and to move
the atoms
%to
towards each other for
the formation
of the required
interatom bonds of diamond.

%Estimates
The estimates
of transformation energies are based on direct calculations and spectrometric measurements of
%exicted
the excited
states of
%alone atom of carbon
sole carbon atom
and are listed in Ref. \cite{Alt}.

Proceeding from
the above
%mentioned
values of
%a energetic
the energy barrier
%at
in the
recalculation on
%one
a single
atom,
we
%shall receive
get
 the necessary energy  for direct
graphite-diamond transition
%graphite - diamond
\begin{equation}
\epsilon=(20\div 200)~\frac{\mbox{Cal}}{\mbox{mole}}=\frac{(20\div 200)10^3 \cdot
2.6 \cdot 10^{19}}{6 \cdot10^{23}}=(1\div 10)~\mbox{eV}/\mbox{atom}.
\label{eq1}
\end{equation}
%Further, for
For the estimations we
%shall
accept
the value $\epsilon=1$ eV/atom. It is necessary to note, that,
%for example,
e.g.
for
the transition
of the  hexagonal
%nitride of boron - in  nitride boron
boron nitride  to the boron nitride
%structure safelite
in the satellite (cubic) or wurcite
%,
structure,
%of
the value $\epsilon$
%will
should be several times
% less
smaller.
Decreasing
%of
$\epsilon$ is possible  also using the known catalytic agents (iron, nickel,
the transition metals of the eighth group of
the periodic table, and also chromium, manganese,
%a
tantalum, etc.)

For estimates we
%shall use
use the
parameters of
the Yerevan Physics Institute electron accelerator
LEA 5,
i.e., the  electron energy  $5$ MeV,
%a
the pulse duration $5$ $\mu$s, a
the pulse current
%in an impulse
$0.75$ A,
the beam diameter
%on an
at the output
$0.2$ cm.

% Run
The run of electrons ($5$ MeV) in graphite  ($Z=6$) is mainly
determined by
the ionization losses and amounts
to $l=1.3$ cm \cite{MKRE}.
During
%all
the entire path length
%an
the electron energy
losses (
%on unity of a trajectory practically are in
practically being within the limits
of 10-20\% on a unit trajectory), are
constant, hence, it is possible to
%consider,
assume
that
the heat  is
released uniformly in a core of
the length $l$.
In the cylinder of length $l$ and
diameter $D$,
%will be contained
 $n$ atoms of graphite  will be contained
\begin{equation}
n=\frac{\pi D^2}{4} l \rho \frac{N_A}{\mu}=2.4 \cdot10^{21},
\label{eq2}
\end{equation}
where $\rho=2.2$ g/cm$^3$
is the graphite density, $N_A=6.23\cdot
10^{23}$ mole$^{-1}$
is the Avogadro number, $\mu=12$ g/mole
is the graphite
%mole
molar mass.
The required energy for transition of all graphite, contained in the cylinder, into
diamond  is
\begin{equation}
W_1=\epsilon\times n=2.4 \times 10^{21}{\mbox { eV}}.
\label{eq3}
\end{equation}
The irradiation with frequency $200$ Hz within one second will provide
the transmission of energy
\begin{equation}
W_2=200\times 5\times 10^6{\mbox {eV}}\frac{0.75 \times 5
\cdot10^{-6}\mbox{(Coulomb)}}{1.6 \cdot10^{-19}\mbox{(Coulomb)}}=2.4\times10^{24}{\mbox { eV}},
\label{eq4}
\end{equation}  that
%on the three order is more, than
is by three orders of magnitude greater than the
energy $W_1$
from Eq. (\ref{eq3}),
necessary for
the transition of the chosen cylinder of graphite into diamond.

%%%%%%%%%%%%%%%%%%%%%%%%%%%%%%%%%%%%%%%%%%%%%%%%%%%%%%%%%%%%%%%%%%%%%%%%%%%
\section{Temperature calculation}\label{sec2}
%%%%%%%%%%%%%%%%%%%%%%%%%%%%%%%%%%%%%%%%%%%%%%%%%%%%%%%%%%%%%%%%%%%%%%%%%
In the above estimates the heat transfer of energy in
%a
the surrounding medium is not taken into account. This
%account
requires the solution of a problem of a thermal conduction with a
%pulse - periodic
pulse-periodic
energy source.

Such problem is solved in  Ref. \cite{Yeger},
for the case when
the released energy of a heat source is constant in time. In the present paper the problem of the composite cylinder with a pulse-periodic energy source is solved.

Let us consider the following problem.
%Region
The region $0<r<a$ (in the cylindrical coordinates) contains a material with thermal
coefficients $k_1, \sigma_1$, and
the region
$r>a$ -- has the parameters $k_2, \sigma_2$, where $k_1$, $k_2$ and
$\sigma_1$, $\sigma_2$ are the heat conductivities and
the temperature conductivitis, respectively.
%correspondingly.

In both regions the initial
 temperatures are  $0^\circ$ C.
 At $t>0$ in
the region $0<r<a$
%in unite time,  in unite volume releases heat quantity
the released heat per unit time and per unit volume is
 $A_0f(t)\,\,~~~$ ( see Fig. 1), where
\begin{eqnarray}
f(t)&=&0,\,\,\,\,\,\,\,\,\,\,\,\,\,t<0,\\
\nonumber
f(t)&=&1,\,\,\,\,\,\,\,\,\,\,\,\,\,n\,t_{b}<t<n\,t_{b}+t_{i},\\
\nonumber
f(t)&=&0,\,\,\,\,\,\,\,\,\,\,\,\,n\,t_{b}+t_{i}<t<(n+1)t_{b},
\label{eq5}
\end{eqnarray}
%where
 and $A_0=Q/(t_{i}\pi a^2 l)=I\epsilon/(\pi a^2 l)$. $Q=I\,t_{i}\,\epsilon$
 is the energy in
 %one
 a single pulse, $I$~ is the beam pulse current, $\epsilon$ is the beam electron energy,
 $l$ is the electrons penetration depth in
 graphite.
\begin{figure}[htb]
\centering
\includegraphics[width=0.99\textwidth]{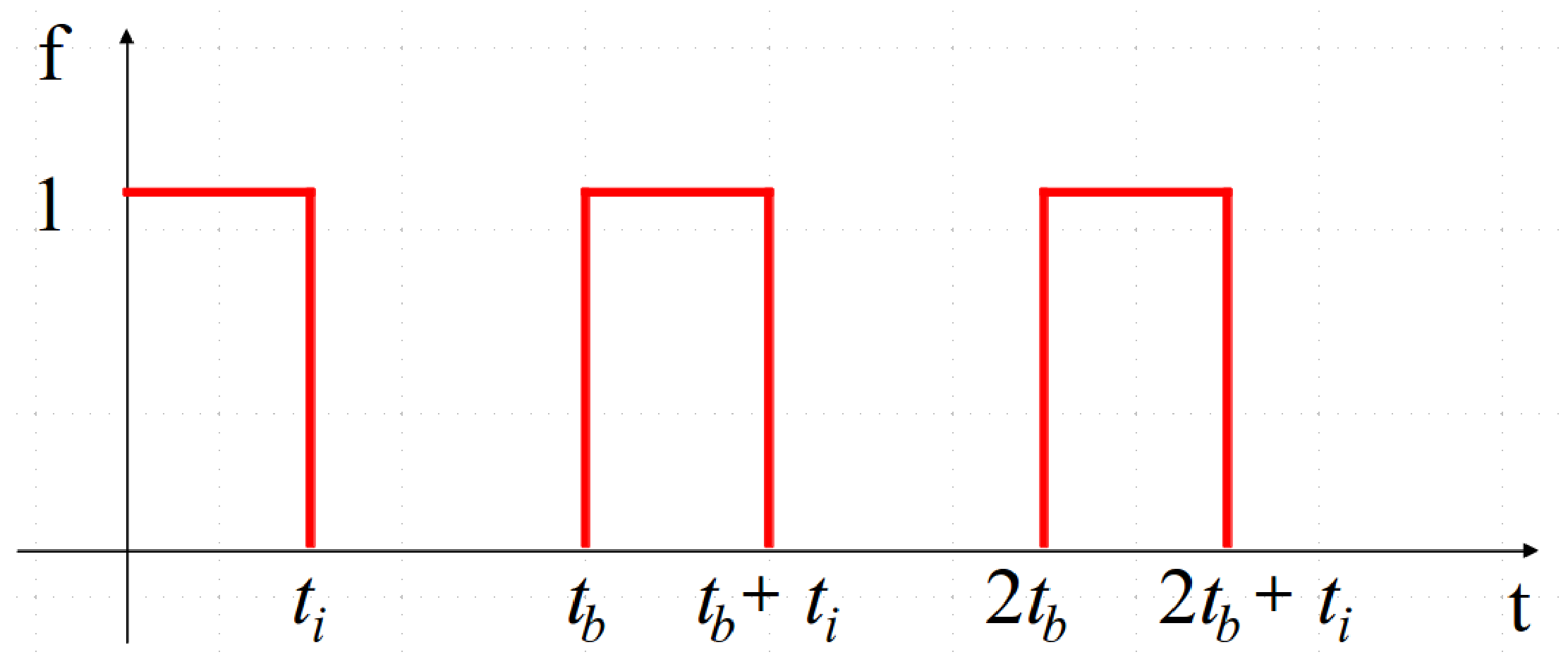}
\caption{The function $f(t)$}
\end{figure}

Let us
%designate through
denote by
$v_1$ and $v_2$
the temperatures in both areas, then the equation of
%a
thermal conduction for these areas
%will be
has the form
\begin{eqnarray}
\frac{\partial^2 v_1}{\partial r^2}+\frac{1}{r}\frac{\partial
v_1}{\partial r}-\frac{1}{\sigma_1}\frac{\partial v_1} {\partial
t}&=&-\frac{A_0}{k_1}f(t),\,\,\,\,\, \,\,\,\,\,0<r<a;
\label{6}\\
 \frac{\partial^2 v_2}{\partial
r^2}+\frac{1}{r}\frac{\partial v_2}{\partial
r}-\frac{1}{\sigma_2}\frac{\partial v_2}{\partial
t}&=&0,\,\,\,\,\,\,\,\,\,\,\,\,\,\,\,\,\,\,\,\,\,\,\,\,\,\,\,\,\,\,\,\,\,\,\,\,\,\,\,\,r>a.
\label{eq7}
\end{eqnarray}
With
the boundary conditions, that the temperature and
%a stream of heat
the heat flow
%on a
 at the boundary
 %of
 between two
 %mediums
 media
 with different properties are continuous
\begin{equation}
v_1=v_2,\,\,\,\,\,\,\,\,\,\, \mbox{and}\,\,\,\,\,\,\,\,\,\,\,\,\,\,\,
k_1\frac{\partial v_1}{\partial r}=k_2\frac{\partial v_2}{\partial r}\,\,\,\,\, \mbox{at}\,\,\,\,\, r=a.
\label{eq8}
\end{equation}

Let us solve these equations with the help of  Laplace transformations ($t %\mapsto
\Leftrightarrow p$).
After
the transformation ($\bar{v}_1, \bar{v}_2$ and
$\bar{f}$
denoting the conversed quantities) they
%become
take the form
\begin{eqnarray}
\frac{\partial^2 \bar{v}_1}{\partial
r^2}+\frac{1}{r}\frac{\partial \bar{v}_1}{\partial
r}-\frac{1}{\sigma_1} \frac{\partial \bar{v}_1}{\partial
r}&=&-\frac{A_0}{k_1}\bar{f}(p)
%\frac{1-e^{-PT_1}}{1-e^{-PT}}
,\,\,\,\,\,0<r<a;
\label{9}\\
\frac{\partial^2 \bar{v}_2}{\partial
r^2}+\frac{1}{r}\frac{\partial \bar{v}_2}{\partial
r}-\frac{1}{\sigma_2} \frac{\partial \bar{v}_2}{\partial
r}&=&0,\,\,\,\,\,\,\,\,\,\,\,\,\,\,\,\,\,\,\,\,\,\,\,\,\,\,\,\,\,\,\,
\,\,\,\,\,\,\,r>a;
\label{eq10}
\end{eqnarray}
%\begin{equation}
%\bar{v}_1&=&\bar{v}_2,\,\,\,\,\,\,\,\,\,\, \and \,\,\,\,\,\,\,\,\,\,\,\,\,\,\, k_1\frac{\partial\bar{v}_1}{\partial r}=k_2\frac{\partial \bar{v}_2}{\partial r}\,\,\,\,\,\,\,\,\,\,\at\,\,\,\,\,\,\,\,\,\, r=a.
%\label{eq11}
%\end{equation}

%\bar{v}_1 = \bar{v}_2,\,\,\,\,\,\,\,\,\,\, and \,\,\,\,\,\,\,\,\,\,\,\,\,\,
\begin{equation}
\bar{v}_1 = \bar{v}_2\,\,\,\,\,\,\,\,\,\, \mbox{and} \,\,\,\,\,\,\,\,\,\,\,\,\,\, k_1\frac{\partial\bar{v}_1}{\partial r}=k_2\frac{\partial \bar{v}_2}{\partial r}\,\,\,\,\,\,\,\,\,\, \mbox{at} \,\,\,\,\,\,\,\,\,\, r=a.
\label{eq11}
\end{equation}
Here  $\bar{f}(p)=\displaystyle\frac{1}{p}\frac{1-e^{-pt_i}}{1-e^{-pt_b}}$,  and  $p$ is the Laplace variable.

%Solutions
The solutions
should be found from the requirements, that at $r=0$, $v_1$ has
the finite value, and at $r\to \infty$ quantity $v_2$ is limited.
%Required
The required
solutions look like
\begin{eqnarray}
\bar{v}_1&=&\frac{1}
%{p^2}
{p}
\frac{\sigma_1 A_0}{k_1}\left(1-\frac{k_2 \sigma_1^{1/2}}{\Delta}K_1(a_2a)I_0(a_1r)\right)
%\frac{1-e^{-pt_i}}{1-e^{-pt_b}}=\frac{\bar{v}_1(p)}{p},
\bar{f}(p)=\bar{v}_1(p)\bar{f}(p)
\label{eq12}\\
\bar{v}_2&=&\frac{1}
%{p^2}
{p}\frac{\sigma_1 A_0}{k_1}
\frac{k_1 \sigma_2^{1/2}}{\Delta}
I_{1}(a_1a)K_0(a_2r)
%\frac{1-e^{-pt_i}}{1-e^{-pt_b}}=\frac{\bar{v}_2(p)}{p},
\bar{f}(p)=\bar{v}_2(p)\bar{f}(p)
\label{eq13}
\end{eqnarray}
where $I_0(x)$, $I_1(x)$, $K_0(x)$, $K_1(x)$ are
the modified Bessel functions,
\begin{eqnarray}
\Delta&=&k_1 \sigma_2^{1/2}\left(I_1(a_1a)K_0(a_2a)+\frac{\sigma}{k}I_0(a_1a)K_1(a_2a)\right),\\
\nonumber
\sigma&=&\sqrt{\frac{\sigma_1}{\sigma_2}},\,\,\,\,\,\,k=\frac{k_1}{k_2},\,\,\,\,\,\,
a_{1,2}=\sqrt{\frac{p}{\sigma_{1,2}}}.
\label{eq14}
\end{eqnarray}

Let us apply the
%theorem of the inversion ?
inversion theorem
\cite{Abr} to $\bar{v}_1$
\begin{equation}
v_1(t)=\frac{1}{2\pi \imath}\frac{\sigma_1 A_0}{k_1}\int_{\gamma-\imath\infty}^{\gamma+\imath \infty}
\left(1-\frac{k_2 \sigma_1^{1/2}}{\Delta}K_1(\mu_2 a)I_0(\mu_1 r)\right)
%\frac{1-e^{-\lambda T_1}}{1-e^{-\lambda T}}
\bar{f}(\lambda)
\frac{e^{\lambda t}}
%{\lambda^2}
{\lambda}
d\lambda ,
\label{eq15}
\end{equation}
where $\gamma$ is to be so large that all singularities of $\bar{v}_1(\lambda)$ lie to the left of
%a
the line ($\gamma-\imath \infty, \gamma+\imath \infty$)(see Fig.2).

\begin{figure}[htb]
\centering
\includegraphics[width=12cm]{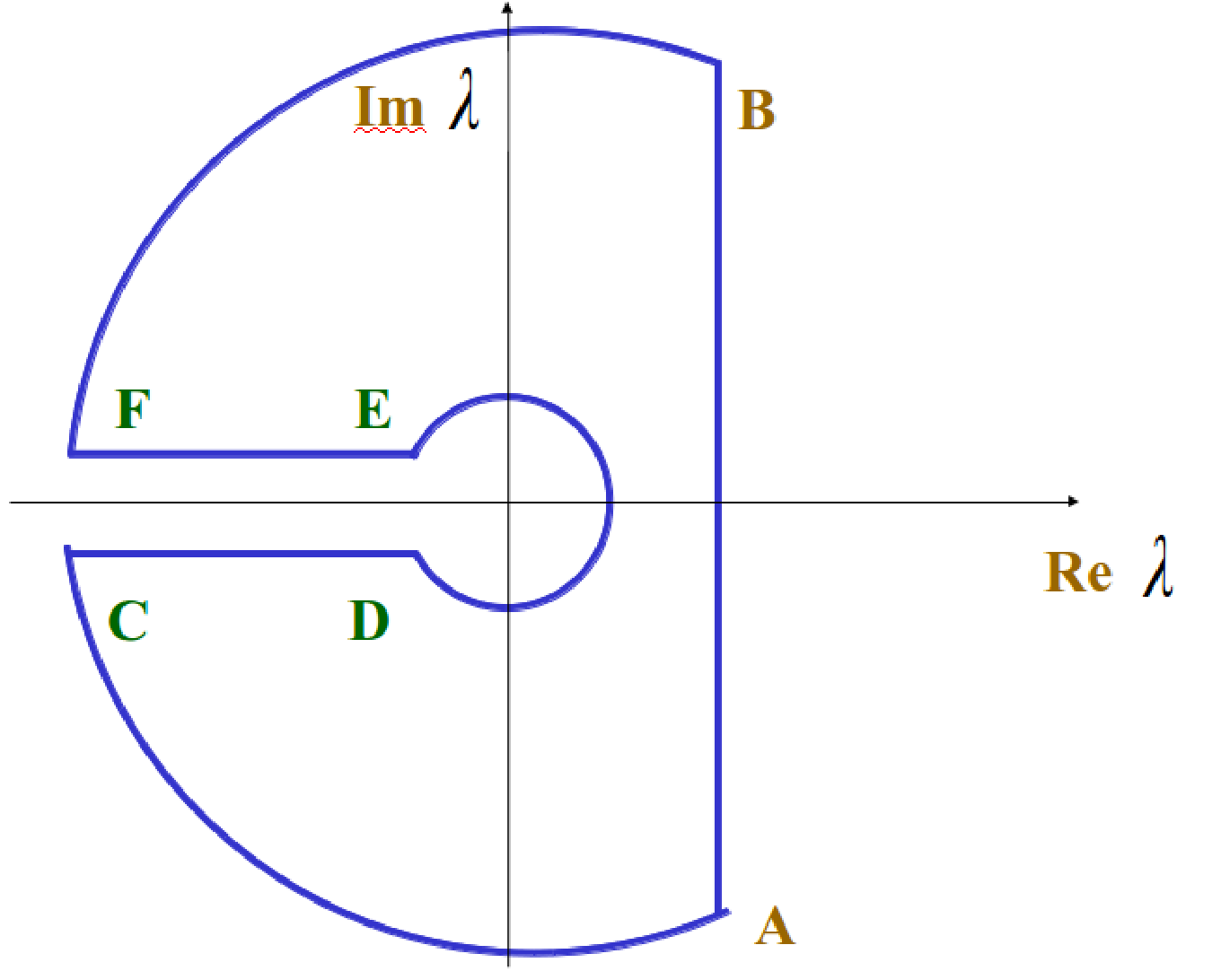}
\caption{The contour of integration}
\end{figure}

In the relation (\ref{eq12}) we have
%exchanged
replaced $p$
%on
with $\lambda$ to
%underline,
emphasize
 that in this relation we consider
 the behavior of
the  function
$\bar{v}_1$, assuming
%its as
it to be
a function of
the complex variable
$\mu_{1,2}=\sqrt{\frac{\lambda}{\sigma_{1,2}}}$ .

The integrand in (\ref{eq15}) has
a  point of
branching at $\lambda=0$ and
% has
simple poles at $\lambda_{n_0}=\pm \imath \displaystyle\frac{2\pi
n_0}{t_{b}}$, where $n_0=1,2,3...$. In this case we use a contour (see
Fig. 2) with
%a slit
a cut along the negative
% material axis
real semiaxis
so that
$\bar{v}(\lambda)$ is a single-valued function $\lambda$ on a
contour and outside
%of
it. The argument $\lambda$ on $EF$ is
$\pi$, and on $CD$  the argument is $-\pi$.

Using the residue formula\,\,
$$\mathop {Res }\limits_{x=a }\frac{\phi(\lambda)}{\psi(\lambda)}=\frac{\phi(a)}{\psi'(a)},\,\,\,\,\,
\phi(a)\neq 0,\,\,\,\,\, \psi(a)=0,\,\,\,\,\, \psi'(a)\neq 0,$$ we
obtain
\begin{equation}
(v_1)_{res}=\frac{\sigma_1 t_{b} a_0}{\pi^2k_1}\sum_{n_0=1}^{\infty}\frac{\sin\frac{\pi n_0t_{b}}{t_i}}{n_0^2}
\eta_{n_0}(r)\sin\left[\frac{2\pi n_0}{t_{b}}\left(t-\frac{t_i}{2}\right)+\epsilon_{n_0}(r)\right],
\label{eq16}
\end{equation}
where functions $\eta_{n_0}(r)$ and $\epsilon_{n_0}(r)$ have the form
\begin{equation}
\eta_{n_0}(r)=\sqrt{Re^2(A+B)+Im^2(A-B)};\,\,\,\,\,\epsilon_{n_0}(r)=\arcsin \frac{A-B}{A+B},
\label{eq17}
\end{equation}

\begin{equation}
A,B=1-\frac{\sigma}{k} \frac{I_0\left(N_1 r e^{\pm \imath\frac{\pi}{4}}\right)K_1\left(N_2 a e^{\pm \imath\frac{\pi}{4}}\right)}
{I_1\left(N_1 a e^{\pm\imath\frac{\pi}{4}}\right)K_0\left(N_2 a e^{\pm \imath\frac{\pi}{4}}\right)+\frac{\sigma}{k}
I_0\left(N_1 a e^{\pm \imath\frac{\pi}{4}}\right)K_1\left(N_2 a e^{\pm \imath\frac{\pi}{4}}\right)},
\label{eq18}
\end{equation}

\begin{equation}
N_{1,2}=\sqrt{\frac{2 \pi n_0}{\sigma_{1,2}t_{b}}}.
\label{eq19}
\end{equation}

Let us  consider now
the integral (\ref{eq15}) (without
%a
the factor $\frac{1}{\lambda}$ or (\ref{eq12}) without $\frac{1}{p}$) on
%a
the contour $ABFEDCA$ at a passage to the limit when the radius of
%a
the large circle $R$ tends to infinity, and the radius
of the  small one
 %-turns
 tends  to zero.
 At $R\to \infty$
 the integral
 %on
 over the  arcs $BF$ and $CA$ tends to zero.
 As the radius of
 %a
 the small circle with
 the center
 %in an
 at the origin of coordinates
 %comes nearer to
 approaches
 zero,
 the integral on this circle also tends to zero.
 At $R\to \infty$ the integral
 %on
 over the $AB$ becomes equal to integral in
 %expression
 Eq. (\ref{eq15}). On
 %a
 the line $EF$ we
 %accept
 assume
 $\lambda=\sigma_1 u^2 e^{\imath \pi}$, then the integral in (\ref{eq15}) will be
\begin{eqnarray}
2\int_0^{\infty}\frac{e^{-\sigma_1 u^2 t}}{u}\left[1-\frac{k_2
\sigma_1^{1/2}}{\Delta} K_1\left(u a
e^{\imath\frac{\pi}{2}}\right)I_0\left(\sigma u r
e^{\imath\frac{\pi}{2}}\right)\right] \frac{1-e^{\sigma_1 u^2
t_i}}{1-e^{\sigma_1 u^2 t_{b}}}du
\nonumber\\
=\frac{2}{k_1\sigma_2^{1/2}}\imath\int_0^{\infty}\frac{e^{-\sigma_1 u^2
t}}{u}\frac{k_2 \sigma_1^{1/2}}{u} \frac{J_0(u r)[J_1(\sigma u
a)\phi-Y_1(\sigma u a)\psi]}{\phi^2+\psi^2} \frac{1-e^{\sigma_1
u^2 t_i}}{1-e^{\sigma_1 u^2 t_{b}}}du.
\label{eq20}
\end{eqnarray}
The integral
%on
over $CD$ gives an expression conjugate
%with
to (\ref{eq20}), with the negative  sign. Summarizing these results,
we obtain for
the integral
\begin{equation}
v_1'(t)=\frac{1}{2\pi}\frac{A_0\sigma_1}{k_1}\frac{\sigma}{k}
\int_0^{\infty}\frac{e^{-\frac{t}{\tau}x^2}}{x}
\frac{J_0(x\frac{r}{a})[J_1(\sigma x)\phi-Y_1(\sigma
x)\psi]}{\psi^2+\phi^2}\frac{1-e^{-\frac{t_i}{\tau}x^2}}
{1-e^{-\frac{t_{b}}{\tau}x^2} }dx,
\label{eq21}
\end{equation}
where parameter $\tau$, functions $\phi$ and $\psi$ have the form
\begin{eqnarray}
\phi&=&J_1(x)Y_0(\sigma x)-\frac{\sigma}{k}J_0(x)Y_1(\sigma x),\\
\nonumber
\psi&=&J_1(x)Y_0(\sigma
x)-\frac{\sigma}{k}J_0(x)J_1(\sigma
x),\,\,\,\,\,\,\,\,\,\,\tau=\frac{a^2}{\sigma_1}.
\label{22}
\end{eqnarray}
%At deriving (\ref{eq21}) relations (\cite{Abr}) have been used
In the derivation of Eq. (21) the following relations have been used [11]:
\begin{eqnarray}
I_1(\pm \imath z)&=&\pm \imath J_1(z),\,\,\,\,\,\,\,\,\,\,I_0(\pm \imath z)=J_0(z),\\
\nonumber
K_0(\pm \imath z)&=&\mp \frac{\imath \pi}{2}\left[J_0(z)\mp \imath Y_0(z)\right],\,\,\,\,\,
K_1(\pm \imath z)=-\frac{\pi}{2}\left[J_1(z)\mp \imath Y_1(z)\right].
\label{eq23}
\end{eqnarray}
The expression in square brackets in the numerator of Eq. (\ref{eq21}) is equal
to $\left[...\right]=\frac{2}{\pi \sigma x}J_1(x)$, where the formula
$$J_0(z)Y_1(z)-J_1(z)Y_0(z)=-\frac{2}{\pi z}$$
 (see, e.g.,\cite{Ryzhik}) is used.
%In view of it, we shall receive
Then we arrive at the expression
\begin{equation}
v_1'(t)=\frac{4}{\pi^2 k}\frac{A_0\sigma_1}{k_1}\int_0^{\infty}\frac{e^{-\frac{t}{\tau}x^2}}{x^2}
\frac{J_1(x)J_0\left(\frac{x}{a}\right)}{\phi^2+\psi^2}
\frac{1-e^{-\frac{t_i}{\tau}x^2}}
{1-e^{-\frac{t_{b}}{\tau}x^2} }dx.
\label{eq24}
\end{equation}
Using the formula \cite{Lavr}
$$\frac{\bar{v}_1'(p)}{p}\Leftrightarrow\int_0^t\bar{v}_1'(\tau)d\tau,$$
%it is finally obtained
we  finally get
\begin{equation}
(v_1(t))_{contour}=\frac{4A_0a^2}{\pi^2kk_1}\int_0^{\infty}\frac{1-e^{-\frac{t}{\tau}x^2}}{x^4}
\frac{1-e^{-\frac{t_i}{\tau}x^2}}
{1-e^{-\frac{t_{b}}{\tau}x^2} }\frac{J_1(x)J_0\left(x\frac{r}{a}\right)}{\phi^2+\psi^2}dx.
\label{eq25}
\end{equation}
%
%General expression for $v(t)=(v(t))_{contour}+(v(t))_{res}$ in
%view of (\ref{eq12}) and (\ref{eq16}) will be
%
With Eqs. (12) and (16) taken into account the general expression for $v(t)=(v(t))_{contour}+(v(t))_{res}$ will take the form
\begin{eqnarray}
v_1(t)&=&\frac{4A_0a^2}{\pi^2kk_1}\int_0^{\infty}\frac{1-e^{-\frac{t}{\tau}x^2}}{x^4}
\frac{1-e^{-\frac{t_i}{\tau}x^2}}
{1-e^{-\frac{t_{b}}{\tau}x^2} }\frac{J_1(x)J_0\left(x\frac{r}{a}\right)}{\phi^2+\psi^2}dx\\
\nonumber
&+&\frac{\sigma_1 t_{b} a_0}{\pi^2k_1}\sum_{n_0=1}^{\infty}\frac{\sin\frac{\pi n_0t_{b}}{t_i}}{n_0^2}
\eta_{n_0}(r)\sin\left[\frac{2\pi n_0}{t_{b}}\left(t-\frac{t_i}{2}\right)+\epsilon_{n_0}(r)\right].
\label{eq26}
\end{eqnarray}

The first term $(v_1)_{res}$ obtained
%of
from the residues
%concerning
at the poles $\lambda_{n_0}=\pm \imath\frac{2\pi n_0}{t_{b}}$, $n_0=1,2,3,...$ represents a part of the solution relevant to a stationary state. It is easy to show, that at small times
\begin{equation}
(v_1(t))_{contour}=\frac{t_i}{t_{b}}\frac{\sigma_1A_0}{k_1}t,
\label{eq27}
\end{equation}
and at large times
\begin{equation}
(v_1(t))_{contour}=\frac{t_i}{t_{b}}\frac{a^2A_0}{2k_1}\ln \frac{4\sigma_2 t}{Ca^2},
\label{eq28}
\end{equation}
where $C=1.7811={e}^\gamma$, $\gamma=0.5772$  is the Euler constant.

%By
In the same way, one can obtain from (\ref{eq7})
\begin{eqnarray}
v_2(t)\!\!&=&\!\!\frac{2A_0a^2}{\pi k_1}\int_0^{\infty}\frac{1-e^{-\frac{t}{\tau}x^2}}{x^4} \frac{1-e^{-\frac{t_i}{\tau}x^2}}{1-e^{-\frac{t_{b}} {\tau}x^2} }\frac{J_1(x)\left[J_0\left(\sigma x\frac{r}{a}\right)\phi - Y_0\left(\sigma x\frac{r}{a}\right)\psi\right]}{\phi^2+\psi^2}dx \qquad\;\;\;\nonumber\\
{}\vphantom{2ex}\nonumber\\
&\,&\qquad\quad\;+\;\;\frac{\sigma_1 t_{b} a_0}{\pi^2k_1}\sum_{n_0=1}^{\infty}\frac{\sin\frac{\pi n_0t_{b}}{t_i}}{n_0^2}\xi_{n_0}(r)\sin\left[\frac{2\pi n_0}{t_{b}}\left(t-\frac{t_i}{2}\right)+\zeta_{n_0}(r)\right],
\label{eq29}
\end{eqnarray}
where functions $\xi_{n_0}(r)$ and $\zeta_{n_0}(r)$ have the form
\begin{equation}
\xi_{n_0}(r)=\sqrt{Re^2(a+b)+Im^2(a-b)};\,\,\,\,\,\zeta_{n_0}(r)=\arcsin \frac{Im(a-b)}{Re(a+b)},
\label{eq30}
\end{equation}

\begin{equation}
a,b= \frac{I_1\left(N_1 a e^{\pm \imath\frac{\pi}{4}}\right)K_0\left(N_2 r e^{\pm \imath\frac{\pi}{4}}\right)}
{I_2\left(N_1 a e^{\pm \imath\frac{\pi}{4}}\right)K_0\left(N_2 a e^{\pm \imath\frac{\pi}{4}}\right)+\frac{\sigma}{k}
I_0\left(N_1 a e^{\pm \imath\frac{\pi}{4}}\right)K_1\left(N_2 a e^{\pm \imath\frac{\pi}{4}}\right)},
\label{eq31}
\end{equation}
and $N_{1,2}$ is given by (\ref{eq19}).
%\begin{equation}
%N_{1,2}=\sqrt{\frac{2 \pi n_0}{\sigma_{1,2}T}}
%\label{eq32}
%\end{equation}

%In figures dependence $v_1(t)$ for a case is given, when the graphite cylinder %is inside sand (Fig.3) and steel (Fig.4).

%For an estimation incipient at an irradiation of graphite of pressures engaging %to viewing an equation of state of solid bodies is necessary

%%%%%%%%%%%%%%%%%%%%%%%%%%%%%%%%%%%%%%%%%%%%%%%%%%%%%%%%%%%%%%%%%%%%%%
\begin{figure}[htb]
\centering
\includegraphics[width=12cm]{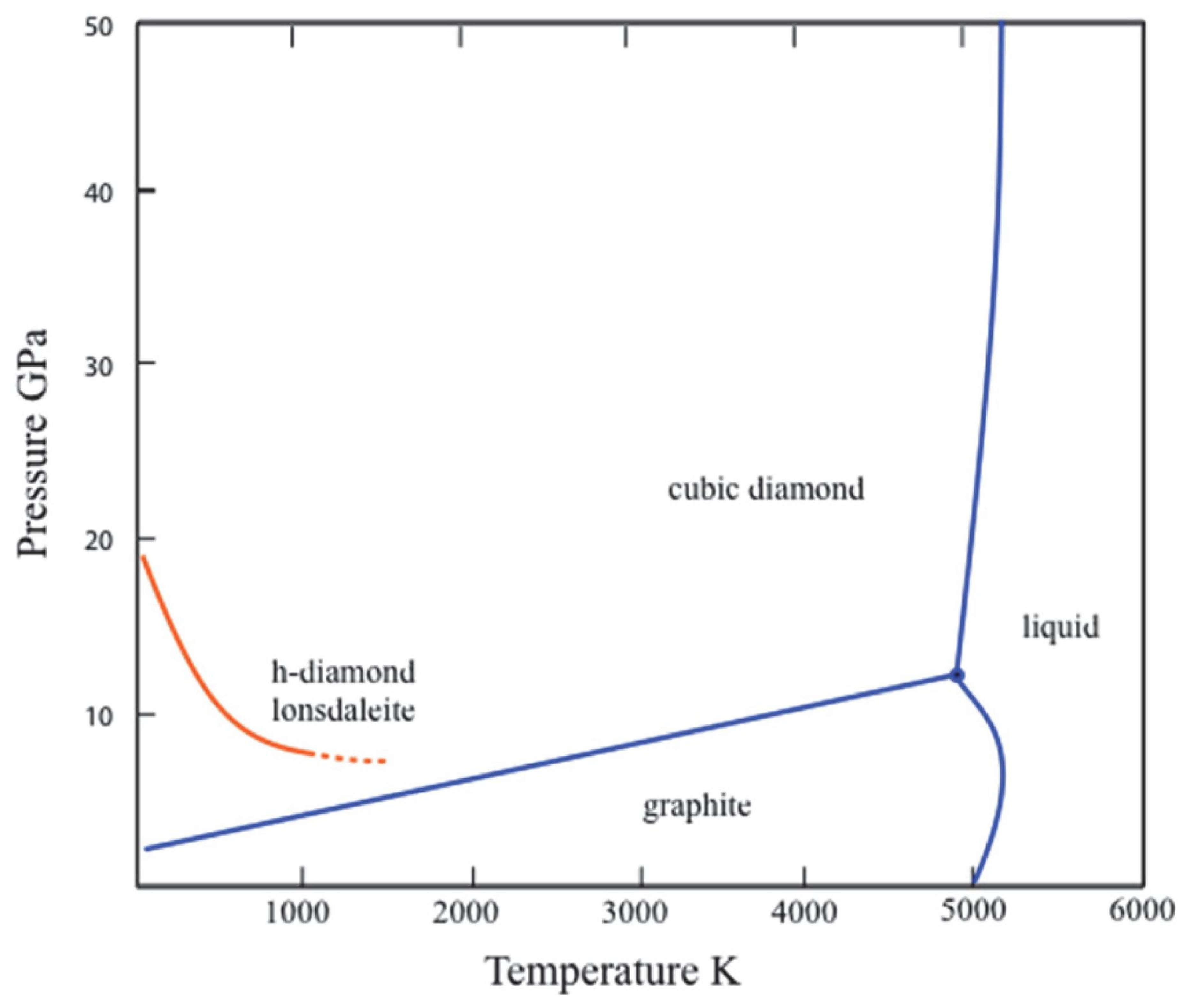}
\caption{Phase diagram of graphite}
\end{figure}

\section{Conclusion}

%%%%%%%%%%%%%%%%%%%%%%%%%%%%%%%%%%%%%%%%%%%%%%%%%%%%%%%%%%%%%%%%%%%%%%

%We shall consider,
It should be taken into account
that for
the temperature available in Yerevan Physics Institute  accelerators   temperature ($T\ll 10000^\circ K$)  the role of electronic terms can be neglected; and the elastic pressure
%is  an order of magnitude close
comparable with the thermal one or exceeds
%thermal.
it.
 Therefore, for an estimate of
 %quantities
 the values of pressures and temperatures, it is possible to
%be restricted only by
 restrict ourselves to the
 thermal terms only ($T\gg T_0$,
  $T_0$ being the  room temperature)
\begin{equation}
P=H\frac{E}{V},\,\,\,\,\,\,\,\,\,\,E=3NkT,
\label{eq33}
\end{equation}
where $P$ is the pressure, $H$ is the Hrunaisen coefficient, which we take
be equal to 2 for graphite, $V$ is the graphite volume, $N$ is the number of atoms in the volume $V$,
$k=1.38\times10^{-23}$ J/K  is the Boltzmann constant, $T$ is the temperature.
Using
%these equations
Eqs. (\ref{eq25}) -- (\ref{eq33}) we find, that the temperature $T=3000$$^\circ$ C corresponds to pressure $P=27\,$ GPa
in the case of sand
%surrounding
environment,  and
the temperature $T=1000^\circ$ C corresponds to $P=9$ GPa in the case of
%steal surrounding.
steel environment

These parameters $(3000^\circ$ C, 27 GPa), $(1000^\circ$ C, 9 GPa)
% on
in the ($P,T$) diagram figure a point in the field of stability of diamond (see Fig.3).
 We shall note also, that direct equations of state $P=9.1\times(10^{-3}T)$ GPa
 %on
 in the
 ($P,T$) diagram lays in the field of stability of diamond
 %, is
 which is higher than
 %a
 the curve of equilibrium diamond - graphite.

%%%%%%%%%%%%%%%%%%%%%%%%%%%%%%%%%%%%%%%%%%%%%%%%%%%%%%%%%%%%%%%%%%%%%%%
\section{Acknowlegments}
%%%%%%%%%%%%%%%%%%%%%%%%%%%%%%%%%%%%%%%%%%%%%%%%%%%%%%%%%%%%%%%%%%%%%%%%%
Authors thank to Dr. A.S. Ayriyan for useful discussion and remarks.
For KBO and NShI this work was supported by the Science Committee of the
Ministry of Science and Education of the Republic of Armenia
(grant number 13-1C080).
KBO thanks LIT JINR for hospitality and support during his visit.

\end{document}